# Effect of the Fe substitution in Ti-Ni shape memory alloys

**T.P. Yadav [1], Durgesh K. Rai[1,2], V.S. Subrahmanyam[1] and O.N.Srivastava[1]**

[1]Department of Physics, Banaras Hindu University, Varanasi-221005, India
[2] Material Science Programme, IIT Kanpur, India

## ABSTRACT

Shape memory Ti-Ni alloys attracted much attention in the recent years, since they are shape memory, intelligent as well as functional materials. In the present investigation $Ti_{51}Ni_{49}$ and $Ti_{51}Ni_{45}Fe_4$ alloys were synthesized through radio frequency (RF) induction melting. The alloy was characterized through x-ray diffraction (XRD), scanning electron microscopy (SEM) , Mössbauer spectroscopy (MS) and positron annihilation techniques (PAT).The Fe substitution stabilized the TiNi type cubic (a=2.998 Å) phase. The surface microstructure and presence of the oxide layer in $Ti_{51}Ni_{45}Fe_4$ alloy have been investigated by SEM. The Positron annihilation measurements indicated a similar bulk electron density in both the as-cast and annealed (1000 °C for 30 hrs) alloys, typically like that of bulk Ti. Mössbauer spectroscopy studies of as-cast and annealed iron substituted samples showed regions in the samples where nuclear Zeeman splitting of Fe levels occurred and an oxide phase was found to be present in as cast $Ti_{51}Ni_{45}Fe_4$ alloy, while annealed sample indicated the presence of bcc iron phase .

Key words: shape memory alloys, Mössbauer spectroscopy, positron annihilation Techniques, nuclear Zeeman splitting

Corresponding author:Tel.: +91-542-2307308, Fax:+91 542 2368468.

E-mail address: yadavtp22@rediffmail.com(T.P.Yadav) , vsspt@yahoo.com(VSS)

# 1. Introduction

Shape memory alloys (SMA's) exhibit two very unique properties, pseudo-elasticity, and the shape memory effect . In the last decade, shape memory alloys (SMA) have attracted much interest as functional materials [1, 2, 3]. Ti-Ni-based shape memory alloys have been widely used as actuators, sensors, and medical devices because of their excellent shape memory properties as well as superelasticity, good mechanical strength and ductility, these properties are related to the thermo-elastic phase transformation [4]. Especially, it has been found that certain microstructures form in Ti-Ni alloys cause different shape memory behavior and improve the shape memory characteristics and mechanical properties [5]. The Ti–Ni based alloys are the most important practical shape memory alloys (PSMA) with excellent mechanical properties [6]. There are many phase transformations in Ti–Ni-based alloys system, which include not only diffusion-less / martensitic transformations, from which shape memory and superelastic effects arise, but also diffusional transformations. Thus even the latter transformations have been used effectively to improve shape memory characteristics [7]. Thus the alloy system serves as an excellent case study in physical metallurgy, as is the case for steels where all kinds of phase transformations are utilized to improve the physical properties [8]

      Ternary alloying elements affect the transformation behavior and shape memory characteristics in the recent years. There has been much effort to modify Ti–Ni shape memory alloys by adding various alloying elements to the binary system. It was found alloying elements often alter transformation temperature greatly [9-10]. It is found that most elements lower the transformation temperature; but there are only a few, notably Pd, Pt, Au, Zr and Hf, which increase transformation temperature. Therefore, Ti–Ni–Pd, Ti–Ni–Zr and Ti–Ni–Hf are considered as candidates for high temperature shape memory alloys [11]. The substitution of Cu for Ni in Ti-Ni reduces the composition sensitivity of martensitic transformation temperature $M_s$ as well as transformation hysteresis [12]. Fabrication of Ti–Ni–Cu ribbon by melt-spinning technique has been proved suitable for producing alloys with Cu content up to 25 at.%, which is too brittle if made using

conventional melting and casting [13]. $Ti_{50}Ni_{25}Cu_{25}$ ribbon has been widely studied because of its one-stage B2 - B19 transformation, the small transformation hysteresis, and most attractively, the large transformation strain [14].

Alloying elements not only change transformation temperature, but also often change the product of the transformation or the transformation route. A typical example is that Ti–Ni undergoes a direct B2–B19 transformation; but with the addition of Fe, Ti–Ni–Fe alloy shows a two-stage transformationB2– R–B19 [15], as it can be noticed that with increasing Fe content, B2–R transformation becomes more and more separated from the R–B19 transformation. This feature enables a characterization of R-phase over wide temperature range without being interrupted by the second transformation R–B19. Due to the high stability of B19, there occurs only one stage transformation B2–B19. From elastic constants of Ti–Ni–Fe alloy, it can be found that Fe addition increases C44 appreciably. Since C44 is crucial for B19 formation, the hardening of C44 indicates that B19 is more unstable, thus the B19 moves up compared with binary alloys.

The aim of the present investigation is to synthesize $Ti_{51}Ni_{49}$ and $Ti_{51}Ni_{45}Fe_4$ alloys and to study the effect of the Fe substitution with regard to the phase stability, microstructural change, bulk electron density and presence of the Fe in Ti-Ni shape memory alloys. We have selected the concentration of Fe as the optimum concentration where the phase remains the same as that of $Ti_{51}Ni_{49}$.

## 2. Experimental details

The alloys with composition $Ti_{51}Ni_{49}$ and $Ti_{51}Ni_{45}Fe_4$ have been synthesized by melting stoichiometric mixture of Ti (99.92 %), Ni (99.9%), Fe (99.9%), in a radio frequency induction furnace (18 kW). The individual elements were at first mixed in correct stoichiometric proportions and pressed into a cylindrical pellet of 1.5 cm diameter, 1 cm thickness by applying a pressure of ~2.76 × $10^4$ N/$m^2$. The pellet (5 g by weight) was then placed in a silica tube surrounded by an outer Pyrex glass jacket. Under continuous flow of argon gas into the silica tube, the pellet was melted using a radio frequency induction furnace (18 kW). During melting process, water was circulated in the outer jacket around the silica tube to reduce the contamination of the alloy. For each composition prepared, the melting atmosphere was further purified by previously melting Ti buttons taking advantage of its getter properties with respect to $O_2$. The alloys were melted repeatedly two to three times to achieve homogeneity. Annealing process is generally expected to increase the homogeneity of the alloy with improvement in a single phase of the crystal growth. Keeping this in view, the as-cast ingots were subjected to annealing. At first ingots were placed in a silica tube, flushed two to three times with high purity argon gas and then evacuated to a pressure of ~$1.32 \times 10^{-7}$ atm, finally the tube was sealed. It was then placed in a furnace and annealed at 1000 °C (±10 °C) for time spans ranging from 10 to 30h. It was found that annealing for 30 h produced the optimum material. Therefore, this time period was maintained for all further annealing runs. The gross structural characterization to identify the crystal structure of the alloy was done by employing X-ray diffraction technique. The ingots were crushed, mechanically ground and the powdered sample was subjected to X-ray diffraction studies employing a Philips X-ray powder diffractometer PW-1710 equipped with a graphite monochromator and CuKα radiation, $α = 1.5406$ Å. The XRD patterns of the as-cast and annealed samples were taken. The surface microstructural characteristics of the as cast and annealed samples were monitored by employing Philips XL-20 series scanning electron microscope (SEM) in secondary electron imaging mode. Positron annihilation measurements have been made using a fast coincidence system with a resolution of 300ps. The Mössbouer spectroscopy measurements have been made using a conventional constant acceleration Mössbauer spectrometer. $^{57}$Co in a rhodium matrix was used as the radioactive source for Mössbauer measurements.

## 3. Results and discussion

In order to perform the phase identification of $Ti_{51}Ni_{49}$ and $Ti_{51}Ni_{45}Fe_4$ alloys, the structural characterizations have been carried out employing XRD techniques. The XRD patterns of as cast and annealed sample of the above compositions are shown in fig. 1 (a-c). The crystallographic structure of the alloy under study has been characterized as Ti(Ni,Fe) and $Ti_2$(Ni,Fe) type phases XRD of the as cast $Ti_{51}Ni_{49}$ sample shown in fig.

1a concluded that $Ti_2Ni$ (cubic) was the major phase obtained and the TiNi phase can be enhanced by annealing at a temperature of about 1000 °C. The phase diagram clearly shows that the $Ti_2Ni$ phase would become unstable whereas TiNi phase would remain stable at that temperature. Thus the sample was annealed at 1000 °C for 20 hours after putting in a silica tube and evacuating, in order to obtain a single phasic sample of TiNi (cubic).

In this case TiNi sample had actually gone through a phase transition, so the peaks of TiNi have gone up while that of $Ti_2Ni$ had gone down. The sample was further annealed 30 hours, It was evident from the XRD pattern (Fig. 1c) that a major phase transition has occurred so that we had finally obtained a single phase TiNi type cubic phase in $Ti_{51}Ni_{49}$ alloy. This was followed by preparation of $Ti_{51}Ni_{45}Fe_4$ under similar conditions and from XRD analysis (fig.1 (b-c)) as expected from earlier experience; we got biphasic alloys which had the two phases. The only difference that was evident at first sight was that here the TiNi phase was the major phase as against $Ti_2Ni$ in the previous case. On similar line as before, it was then decided that the sample be annealed at 1000°C for 20 hours. It was quite evident that our material has gone through a major phase transformation from biphasic to a single phasic TiNi (cubic structure) alloy. A comparison of the XRD patterns in fig 1(a) – (b-c) of TiNi and TiNiFe annealed at 1000°C reveals that under similar experimental conditions of preparation and annealing, a better matching of the various peaks and the respective intensities with standard TiNi (a=2.998Å cubic) alloy, as obtained from (JCPDS files [16]) , was found for the iron substituted sample. It should also be mentioned that upon annealing the as-cast alloy, the strain as determined from XRD peaks, decreases from 0.351% to 0.213%. This may be attributed to homogenization effects upon annealing.

Fig. 2(a-d) show the scanning electron micrographs as imaged with secondary electrons for the as-cast and annealed of $Ti_{51}Ni_{49}$ and $Ti_{51}Ni_{45}Fe_4$ samples respectively, which provide important information about the surface morphology of the sample. It is to be mentioned here that the as-prepared sample was very hard. A grinder had to be used and the sample was cut and SEM was taken. Larger view of the sample surface contains

steps, stairs, and cracks at various places. Also, small regions with globular shaped surfaces can be seen. SEM of annealed titanium nickel (Single phasic) alloy shows roughness enhanced crystalline structures, effects of surface oxidation upon thermal treatment. A different region "A" on the same sample surface shows large cracks that show signs of thermal treatment. In larger view of the Titanium Nickel Iron as cast alloy in fig 2 (c), the surface shows, some interesting features like black spots marked easily, which may be the regions of oxidized Fe on the surface, where as in the annealed $Ti_{51}Ni_{45}Fe_4$ alloy in fig. 2d, no dark spots have been seen. It may be due the formation of $TiO_2$ oxide material.

Positron annihilation lifetime measurement is a powerful technique to probe the defect structure of solids [17]. Peak-normalised positron lifetime spectra of the as-cast and annealed TiNi alloy samples have been shown in fig. 3. The lifetime data have been analysed using the Kansy Program [18]. Three components were resolved. The results are presented in Table 1. The smallest lifetime is characteristic of positrons annihilating in the bulk of the alloy. This value in both the samples is similar (within the errors), and is a little higher than that of positron lifetime in titanium bulk, namely, 148 ps [17]. The second lifetime may represent positron annihilations at inhomogeneities, interfaces, and open volume defects, if any, which are known to trap positrons. The long lifetime component indicates the possible presence of positronium (Ps). However, this is unlikely in the bulk of a metal or an alloy, but, Ps could be formed at very large open volume defects, and at surfaces. However, the nature of the two longer components can not be ascertained without further study.

Using Mössbauer spectroscopy technique it is possible to measure the changes in the nuclear energy levels with an accuracy of $\sim 10^{-8}$ eV. This allows a direct determination of hyperfine interactions where the nuclear energy levels are perturbed by the electronic environment around it. The spectra are fitted using least squares method and assuming each spectrum to be a sum of Lorentzian functions. The Mössbauer spectrum clearly shows (fig 4 a, b) the overlapping sextets. Two sextets with HMF values, namely, 289.0 kOe, with 43% intensity and 494.1 kOe with 18.6% intensity were resolved (the remaining being singlet and a doublet with the rest of the intensity). The different HMF values for the sextets indicate different types of environments for Fe, under which the nuclear Zeeman splitting of the Fe levels occur. Both of these values are different from that of pure iron (bcc). The higher value may be due to oxidized iron ($Fe_3O_4$). This observation seems to be in agreement with the SEM observation on as-cast iron substituted sample. Interestingly, upon annealing, a single sextet pattern with a HMF, 330.0 kOe, characteristic of that of ferromagnetic phase of pure iron (bcc) emerges, with an intensity of about 16%. This indicates that pockets of ferromagnetic iron phase may be present within the sample. It appears that the $Fe_3O_4$ local regions upon annealing yield to give rise to iron atom clusters. Such developments may be possible while annealing owing to different thermal reactivities of iron and titanium, which is present in the bulk in abundance (51%), towards presence of oxygen. It is possible that the titanium atoms show stronger activities towards the oxygen present in the bulk, leading to formation of titanium oxide and liberation of iron (in pure form) leading to formation of bcc iron phase. The very fact that the amount of iron in the sample is only 4% and that of its oxide even an order less, itself makes it impossible to detect any such developments in the bulk by XRD as that would hardly cross the noise levels of the instrument.

**Conclusions:**

We have synthesized $Ti_{51}Ni_{49}$ and $Ti_{51}Ni_{45}Fe_4$ shape memory alloy and studied their structural and microstructural characteristics. The Mössbauer spectroscopy (MS) studies have also been carried out. The basic conclusions emanating from the present studies can be summarized as in the following:

1. The as-cast $Ti_{51}Ni_{49}$ alloys were found to be biphasic, having $Ti_2Ni$ and $TiNi$ phases where as after annealing at 1000 °C for 30 h the $Ti_2Ni$ phase seems to disappear.
2. The $Ti_{51}Ni_{45}Fe_4$ alloy reveals that upon iron substitution the alloy seems to favor TiNi phase. On the other hand, the lattice strain of the annealed $Ti_{51}Ni_{45}Fe_4$ alloy which is 0.213% is lower than that of 0.351% for the as-cast alloy. The decrease of lattice strain is attributed to improvement in homogeneity upon annealing.
3. SEM images revealed the presence of various surface morphological features like striations, steps, stairs, cracks, and globules. Thermal treatment resulted in enhanced surface roughness and large cracks at a few places on the surface. Interesting features like black spots were visible on iron substituted sample.

4. Mössbauer studies of as-cast and annealed iron substituted samples showed regions in the samples where nuclear Zeeman splitting of Fe levels occurred. While the as-cast sample showed two different types of such Fe surroundings (indicated by two sextets), the annealed sample showed a single sextet with a hyperfine magnetic field value characteristic of ferromagnetic phase of iron (bcc).

**Acknowledgements**

The authors like to thank Prof. H.C.Verma, Prof. R.S. Tiwari and Dr.N.K.Mukhopadhyay for Helpful discussions. Financial assistance from MNES is gratefully acknowledged. One of the authors (TPY) is thankful to CSIR for the award of SRF Fellowship.

**References:**

[1] J.D.Busch,A.D.Johnson, C.H.Lee D.A.Stevenson J.Appl. Physics 68(1990)6224

[2] A.Ishida, A.Takei,M.Sato,S.Miyazaki Thin Solid Fils 281-182 ( 1996) 337

[3] A.Ishida, ,M.Sato, T. Kimura, S.Miyazaki  Phil Mag A 80 (2000) 967

[4]  K.Otsuka , X Ren Intermetallics 7 (1999)511

[5]  A.Ishida, M.Sato,S.Miyazaki  Mater Sci Eng A 273~275 ( 1999) 754

[6] J.X. Zhang , M. Sato, A. Ishida Acta Materialia 54 (2006) 1185

[7] Yumei Zhou, Jian Zhang, Genlian Fan,Xiangdong Ding , Jun Sun, Xiaobing Ren , Kazuhiro Otsuka   Acta Materialia 53 (2005) 5365

[8]  K.Otsuka , X Ren   progress in Materials Science 50 (2005) 511

[9]  G.P. Chenga, Z.L. Xie a,., Y. Liub  Journal of Alloys and Compounds 415 (2006) 182.

[10]  DR Angst ,  PE Thoma , MY  Kao    J Phys IV  (ICOMAT-95)  C8 (1995) 747.

[11]  K Otsuka , D  Golberg  . In: Vincenzini, editor. Intelligent  mater and systems. Techna  Srl. (1995)  55.

[12]  ZS Basinski , JW Christian . Acta Metall  2 (1954) 101.

[13] T Honma  In: Funakubo H, editor. Shape memory alloys. New York: Gordon and
     Breach Science  Publishers (1987) 61.

[14]  K Otsuka , T. Kakeshita  MRS Bulletin 27 (2002) 91

[15]   T. Honma , M. Matsumoto ,Y Shugo, M.Nishida, I YamazakiI. In:H. Kimura, O.Izumi,
     editors. Proc 4th Int Conf on Titanium, Kyoto. AIME   (1980) 1455

[16]  Powder Diffraction File, Joint Committee on Powder Diffraction Standards,
     Swarthmore, PA, 1990, No. 60696.

[17] (for example) Positron Solid State Physics, (Eds.) W. Brandt and A. Dupasquier, (1983) Proceedings of the International School of Physics, Enrico Fermi, 1981, Course LXXXIII (North-Holland, Amsterdam).

[18] J. Kansy, Microcomputer program for analysis of positron annihilation lifetime spectra. Nucl. Instrum. Meth. A 374 (1996) 235.

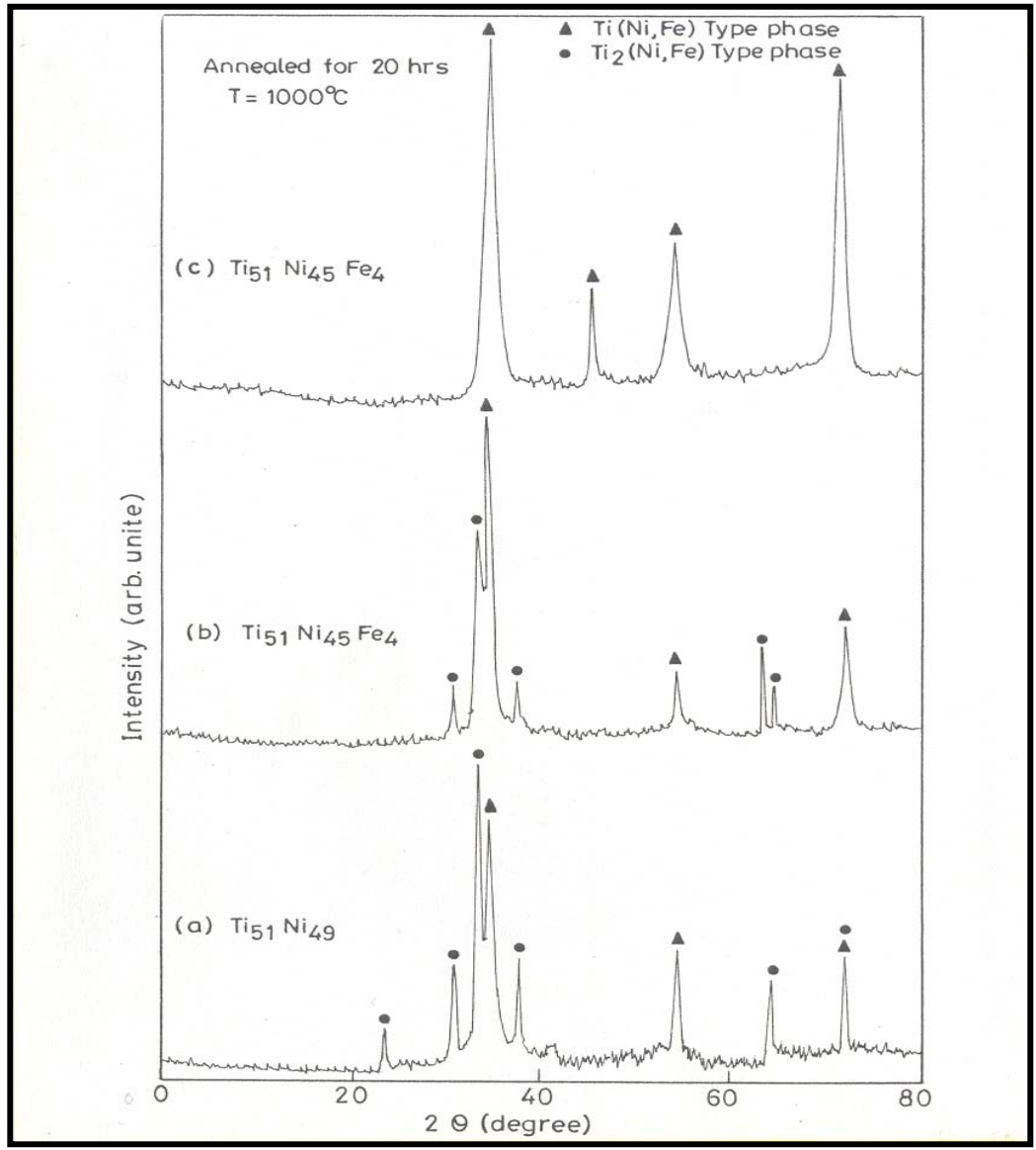

Fig1. X-ray diffraction patterns of (a) Ti 51 Ni49 as cast alloy (b) Ti 51 Ni 45 Fe 4 as cast alloy and (c) Ti 51 Ni 45 Fe 4 alloy annealed at 1000 °C for 20 hours indicating that the formation of Ti NI type phase ,

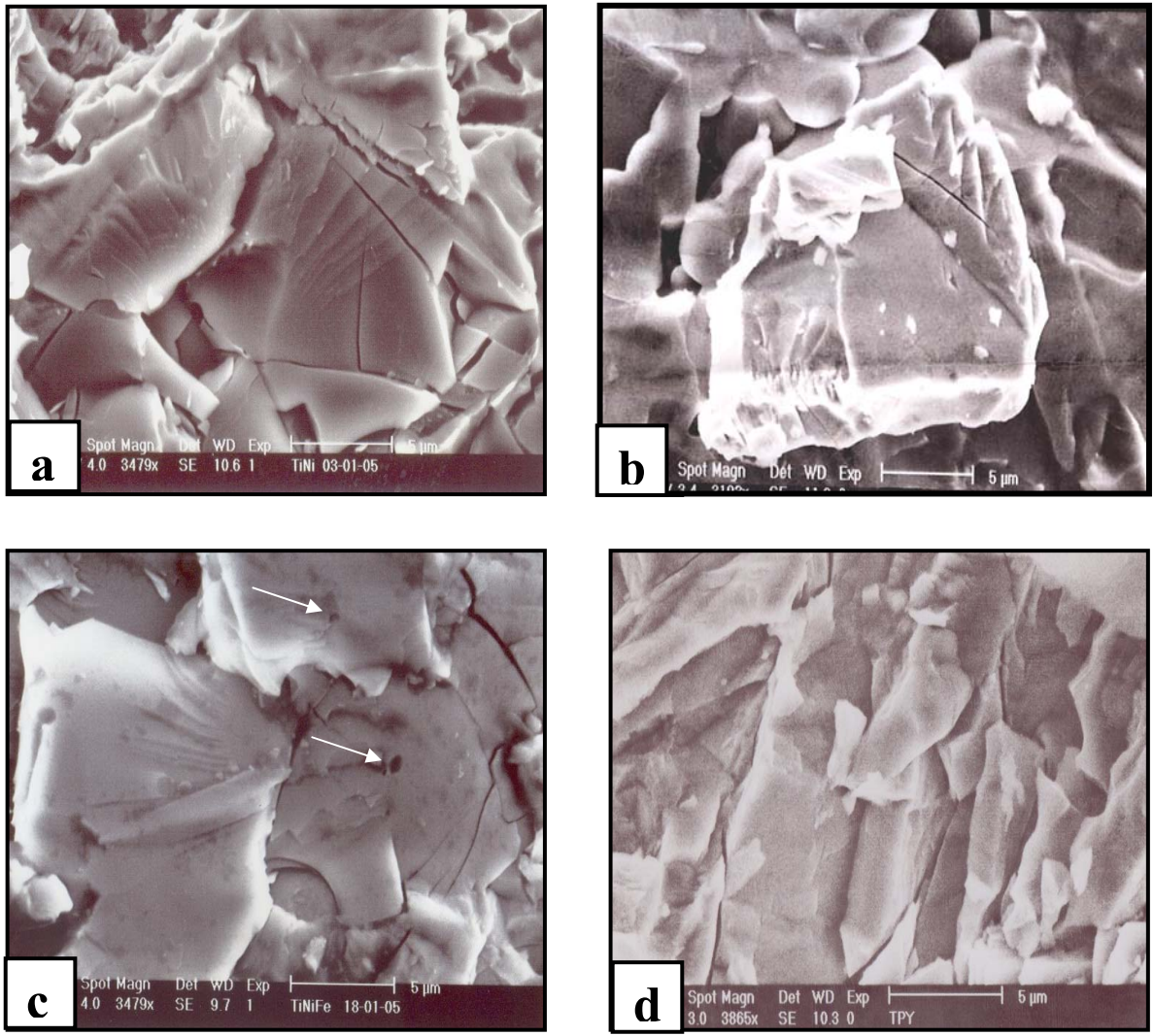

Fig.2 (a) Scanning electron micrograph of the as cast $Ti_{51}Ni_{49}$ alloy. (b) Scanning electron micrograph of the $Ti_{51}Ni_{49}$ annealed alloy. The microstructure reveals the presence of several discrete alloy particles. (c) Scanning electron micrograph of the as cast $Ti_{51}Ni_{45}Fe_4$ alloy (d) Scanning electron micrograph of the $Ti_{51}Ni_{45}Fe_4$ annealed alloy.

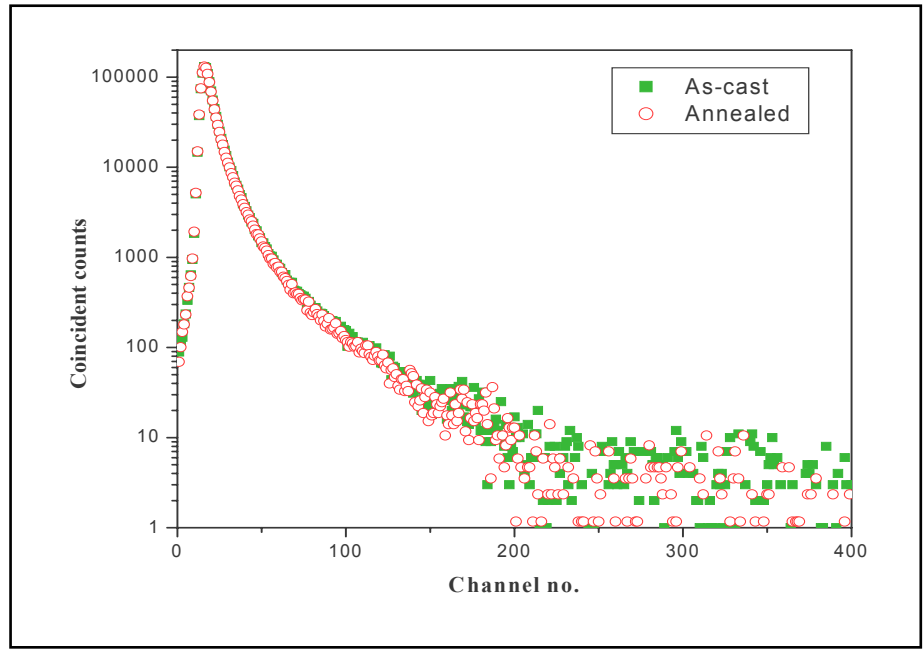

Fig.3. Positron lifetime spectra of the as-cast and annealed TiNi alloy samples.

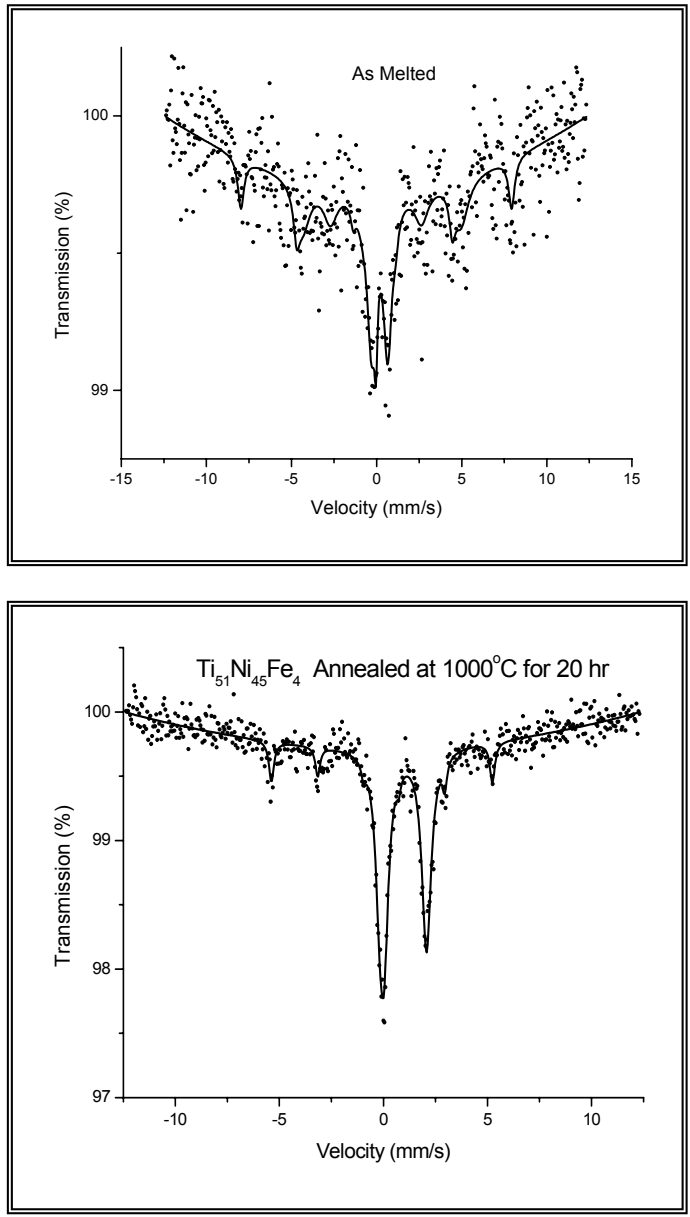

Fig.4.Mössbauer studies of as-cast (a) and annealed iron substituted samples ( b) showed regions in the samples where nuclear Zeeman splitting of Fe levels occurred. While the as-cast sample showed two different types of such Fe surroundings (indicated by two sextets), the annealed sample showed a single sextet with a hyperfine magnetic field value characteristic of ferromagnetic phase of iron (BCC).

Table 1: Positron lifetime parameters in the as-cast and annealed (1000 °C for 30 hrs)

| Alloy | $\tau_1$(ps) | $I_1$(%) | $\tau_2$(ps) | $I_2$(%) | $\tau_3$(ps) | $I_3$(%) |
|---|---|---|---|---|---|---|
| As-cast | 157(2) | 69.0(1.0) | 441(22) | 24.5(1.0) | 1061(14) | 6.1(0.1) |
| 1000 °C for 30 hrs Annealed | 157(2) | 74.0(1.0) | 457(58) | 18.9(1.0) | 910(15) | 7.0(0.1) |